
%
%
%
%
%
%
%
%
\def\standardrisposta{s }\def\reducedrisposta{r }
\def\mplarisposta{mpla }\def\zerorisposta{z }\def\bigrisposta{big }
\def\doublerisposta{d }\def\cartarisposta{e }\def\amsrisposta{y }
\newcount\ingrandimento \newcount\sinnota \newcount\dimnota
\newcount\unoduecol \newdimen\collhsize \newdimen\tothsize
\newdimen\fullhsize \newcount\controllorisposta \sinnota=1
\newskip\infralinea  \global\controllorisposta=0
\immediate\write16 { ********  Welcome to PANDA macros (Plain TeX,
AP, 1991) ******** }
\immediate\write16 { You'll have to answer a few questions in
lowercase.}
\message{>  Do you want it in double-page (d), reduced (r)
or standard format (s) ? }\read-1 to\risposta
\message{>  Do you want it in USA A4 (u) or EUROPEAN A4
(e) paper size ? }\read-1 to\srisposta
\message{>  Do you have AMSFonts 2.0 (math) fonts (y/n) ? }
\read-1 to\arisposta
%
%
%
%
%
\ifx\risposta\standardrisposta \ingrandimento=1200
\message {>> This will come out UNREDUCED << }
\dimnota=2 \unoduecol=1 \global\controllorisposta=1 \fi
\ifx\risposta\bigrisposta \ingrandimento=1440
\message {>> This will come out ENLARGED << }
\dimnota=2 \unoduecol=1 \global\controllorisposta=1 \fi
\ifx\risposta\reducedrisposta \ingrandimento=1095 \dimnota=1
\unoduecol=1  \global\controllorisposta=1
\message {>> This will come out REDUCED << } \fi
\ifx\risposta\doublerisposta \ingrandimento=1000 \dimnota=2
\unoduecol=2  \message {>> You must print this in
LANDSCAPE orientation << } \global\controllorisposta=1 \fi
\ifx\risposta\mplarisposta \ingrandimento=1000 \dimnota=1
\message {>> Mod. Phys. Lett. A format << }
\unoduecol=1 \global\controllorisposta=1 \fi
\ifx\risposta\zerorisposta \ingrandimento=1000 \dimnota=2
\message {>> Zero Magnification format << }
\unoduecol=1 \global\controllorisposta=1 \fi
\ifnum\controllorisposta=0  \ingrandimento=1200
\message {>>> ERROR IN INPUT, I ASSUME STANDARD
UNREDUCED FORMAT <<< }  \dimnota=2 \unoduecol=1 \fi
\magnification=\ingrandimento
%
%
%
%
\newdimen\eucolumnsize \newdimen\eudoublehsize \newdimen\eudoublevsize
\newdimen\uscolumnsize \newdimen\usdoublehsize \newdimen\usdoublevsize
\newdimen\eusinglehsize \newdimen\eusinglevsize \newdimen\ussinglehsize
\newskip\standardbaselineskip \newdimen\ussinglevsize
\newskip\reducedbaselineskip \newskip\doublebaselineskip
\newskip\bigbaselineskip
\eucolumnsize=12.0truecm    
\eudoublehsize=25.5truecm   
\eudoublevsize=6.5truein    
\uscolumnsize=4.4truein     
\usdoublehsize=9.4truein    
\usdoublevsize=6.8truein    
\eusinglehsize=6.3truein    
\eusinglevsize=24truecm     
\ussinglehsize=6.5truein    
\ussinglevsize=8.9truein    
\bigbaselineskip=18pt plus.2pt       
\standardbaselineskip=16pt plus.2pt  
\reducedbaselineskip=14pt plus.2pt   
\doublebaselineskip=12pt plus.2pt    
%
%
\def\Portoffset{}
\def\Landoffset{\hoffset=-.140truein}
\ifx\risposta\mplarisposta \def\Portoffset{\hoffset=1.9truecm
\voffset=1.4truecm} \fi
%
%
\def\Landspec{}
\tolerance=10000
\parskip=0pt plus2pt  \leftskip=0pt \rightskip=0pt
%
%
\ifx\risposta\bigrisposta      \infralinea=\bigbaselineskip \fi
\ifx\risposta\standardrisposta \infralinea=\standardbaselineskip \fi
\ifx\risposta\reducedrisposta  \infralinea=\reducedbaselineskip \fi
\ifx\risposta\doublerisposta   \infralinea=\doublebaselineskip \fi
\ifx\risposta\mplarisposta     \infralinea=13pt \fi
\ifx\risposta\zerorisposta     \infralinea=12pt plus.2pt\fi
\ifnum\controllorisposta=0    \infralinea=\standardbaselineskip \fi
\ifx\risposta\doublerisposta   \Landoffset \else \Portoffset \fi
\ifx\risposta\doublerisposta \ifx\srisposta\cartarisposta
\tothsize=\eudoublehsize \collhsize=\eucolumnsize
\vsize=\eudoublevsize  \else  \tothsize=\usdoublehsize
\collhsize=\uscolumnsize \vsize=\usdoublevsize \fi \else
\ifx\srisposta\cartarisposta \tothsize=\eusinglehsize
\vsize=\eusinglevsize \else  \tothsize=\ussinglehsize
\vsize=\ussinglevsize \fi \collhsize=4.4truein \fi
\ifx\risposta\mplarisposta \tothsize=5.0truein
\vsize=7.8truein \collhsize=4.4truein \fi
%
%
%
%
\newcount\contaeuler \newcount\contacyrill \newcount\contaams
\newcount\contasym
\font\ninerm=cmr9  \font\eightrm=cmr8  \font\sixrm=cmr6
\font\ninei=cmmi9  \font\eighti=cmmi8  \font\sixi=cmmi6
\font\ninesy=cmsy9  \font\eightsy=cmsy8  \font\sixsy=cmsy6
\font\ninebf=cmbx9  \font\eightbf=cmbx8  \font\sixbf=cmbx6
\font\ninett=cmtt9  \font\eighttt=cmtt8  \font\nineit=cmti9
\font\eightit=cmti8 \font\ninesl=cmsl9  \font\eightsl=cmsl8
\skewchar\ninei='177 \skewchar\eighti='177 \skewchar\sixi='177
\skewchar\ninesy='60 \skewchar\eightsy='60 \skewchar\sixsy='60
\hyphenchar\ninett=-1 \hyphenchar\eighttt=-1 \hyphenchar\tentt=-1
\def\bfmath{\cmmib}                 
\font\tencmmib=cmmib10  \newfam\cmmibfam  \skewchar\tencmmib='177
\font\tencmbsy=cmbsy10  \newfam\cmbsyfam  \skewchar\tencmbsy='60
\def\scaps{\cmcsc}                 
\font\tencmcsc=cmcsc10  \newfam\cmcscfam
\ifnum\ingrandimento=1095 
 
\font\bfone=cmbx10 at 10.95pt

\font\capsone=cmcsc10 at 10.95pt 

\else  
 
\font\bfone=cmbx10 at 12pt

\font\capsone=cmcsc10 at 12pt 
\fi
\def\chapterfont#1{\xdef\ttaarr{#1}}
\def\sectionfont#1{\xdef\ppaarr{#1}}
\def\ttaarr{\bf}                
\def\ppaarr{\sl}                

%
%
%
\newfam\eufmfam \newfam\msamfam \newfam\msbmfam \newfam\eufbfam
\def\Loadeulerfonts{\global\contaeuler=1 \ifx\arisposta\amsrisposta
\font\teneufm=eufm10              
\font\eighteufm=eufm8 \font\nineeufm=eufm9 \font\sixeufm=eufm6
\font\seveneufm=eufm7  \font\fiveeufm=eufm5
\font\teneufb=eufb10              
\font\eighteufb=eufb8 \font\nineeufb=eufb9 \font\sixeufb=eufb6
\font\seveneufb=eufb7  \font\fiveeufb=eufb5
\font\teneurm=eurm10              
\font\eighteurm=eurm8 \font\nineeurm=eurm9
\font\teneurb=eurb10              
\font\eighteurb=eurb8 \font\nineeurb=eurb9
\font\teneusm=eusm10              
\font\eighteusm=eusm8 \font\nineeusm=eusm9
\font\teneusb=eusb10              
\font\eighteusb=eusb8 \font\nineeusb=eusb9
\else \def\eufm{\tt} \def\eufb{\tt} \def\eurm{\tt} \def\eurb{\tt}
\def\eusm{\tt} \def\eusb{\tt}    \fi}
\def\loadamsmath{\global\contaams=1 \ifx\arisposta\amsrisposta
\font\tenmsam=msam10 \font\ninemsam=msam9 \font\eightmsam=msam8
\font\sevenmsam=msam7 \font\sixmsam=msam6 \font\fivemsam=msam5
\font\tenmsbm=msbm10 \font\ninemsbm=msbm9 \font\eightmsbm=msbm8
\font\sevenmsbm=msbm7 \font\sixmsbm=msbm6 \font\fivemsbm=msbm5
\else \def\msbm{\bf} \fi \def\Bbb{\msbm} \def\symbl{\msam} \tenpoint}
\def\loadcyrill{\global\contacyrill=1 \ifx\arisposta\amsrisposta
\font\tenwncyr=wncyr10 \font\ninewncyr=wncyr9 \font\eightwncyr=wncyr8
\font\tenwncyb=wncyr10 \font\ninewncyb=wncyr9 \font\eightwncyb=wncyr8
\font\tenwncyi=wncyr10 \font\ninewncyi=wncyr9 \font\eightwncyi=wncyr8
\else \def\cyrill{\sl} \def\cyrilb{\sl} \def\cyrili{\sl} \fi\tenpoint}
\catcode`\@=11
\def\undefine#1{\let#1\undefined}
\def\newsymbol#1#2#3#4#5{\let\next@\relax
 \ifnum#2=\@ne\let\next@\msafam@\else
 \ifnum#2=\tw@\let\next@\msbfam@\fi\fi
 \mathchardef#1="#3\next@#4#5}
\def\mathhexbox@#1#2#3{\relax
 \ifmmode\mathpalette{}{\m@th\mathchar"#1#2#3}%
 \else\leavevmode\hbox{$\m@th\mathchar"#1#2#3$}\fi}
\def\hexnumber@#1{\ifcase#1 0\or 1\or 2\or 3\or 4\or 5\or 6\or 7\or 8\or
9\or A\or B\or C\or D\or E\or F\fi}
\edef\msafam@{\hexnumber@\msamfam}
\edef\msbfam@{\hexnumber@\msbmfam}
\mathchardef\dabar@"0\msafam@39
\catcode`\@=12
\def\loadamssym{\ifx\arisposta\amsrisposta  \ifnum\contaams=1
\global\contasym=1
\catcode`\@=11
\def\dashrightarrow{\mathrel{\dabar@\dabar@\mathchar"0\msafam@4B}}
\def\dashleftarrow{\mathrel{\mathchar"0\msafam@4C\dabar@\dabar@}}
\let\dasharrow\dashrightarrow
\def\ulcorner{\delimiter"4\msafam@70\msafam@70 }
\def\urcorner{\delimiter"5\msafam@71\msafam@71 }
\def\llcorner{\delimiter"4\msafam@78\msafam@78 }
\def\lrcorner{\delimiter"5\msafam@79\msafam@79 }
\def\yen{{\mathhexbox@\msafam@55}}
\def\checkmark{{\mathhexbox@\msafam@58 }}
\def\circledR{{\mathhexbox@\msafam@72 }}
\def\maltese{{\mathhexbox@\msafam@7A }}
\catcode`\@=12
\input amssym.tex     \else
\message{Panda error - First you have to use loadamsmath !!!!} \fi
\else \message{Panda error - You need the AMSFonts for these symbols
!!!!}\fi}
\ifx\arisposta\amsrisposta
\font\sevenex=cmex7               
\font\eightex=cmex8  \font\nineex=cmex9
\font\ninecmmib=cmmib9   \font\eightcmmib=cmmib8
\font\sevencmmib=cmmib7 \font\sixcmmib=cmmib6
\font\fivecmmib=cmmib5   \skewchar\ninecmmib='177
\skewchar\eightcmmib='177  \skewchar\sevencmmib='177
\skewchar\sixcmmib='177   \skewchar\fivecmmib='177
\font\ninecmbsy=cmbsy9    \font\eightcmbsy=cmbsy8
\font\sevencmbsy=cmbsy7  \font\sixcmbsy=cmbsy6
\font\fivecmbsy=cmbsy5   \skewchar\ninecmbsy='60
\skewchar\eightcmbsy='60  \skewchar\sevencmbsy='60
\skewchar\sixcmbsy='60    \skewchar\fivecmbsy='60
\font\ninecmcsc=cmcsc9    \font\eightcmcsc=cmcsc8     \else
\def\cmmib{\fam\cmmibfam\tencmmib}\textfont\cmmibfam=\tencmmib
\scriptfont\cmmibfam=\tencmmib \scriptscriptfont\cmmibfam=\tencmmib
\def\cmbsy{\fam\cmbsyfam\tencmbsy} \textfont\cmbsyfam=\tencmbsy
\scriptfont\cmbsyfam=\tencmbsy \scriptscriptfont\cmbsyfam=\tencmbsy
\scriptfont\cmcscfam=\tencmcsc \scriptscriptfont\cmcscfam=\tencmcsc
\def\cmcsc{\fam\cmcscfam\tencmcsc} \textfont\cmcscfam=\tencmcsc \fi
\catcode`@=11
\newskip\ttglue
\gdef\tenpoint{\def\rm{\fam0\tenrm}
  \textfont0=\tenrm \scriptfont0=\sevenrm \scriptscriptfont0=\fiverm
  \textfont1=\teni \scriptfont1=\seveni \scriptscriptfont1=\fivei
  \textfont2=\tensy \scriptfont2=\sevensy \scriptscriptfont2=\fivesy
  \textfont3=\tenex \scriptfont3=\tenex \scriptscriptfont3=\tenex
  \def\mcal{\fam2 \tensy}  \def\mmit{\fam1 \teni}
  \textfont\itfam=\tenit \def\it{\fam\itfam\tenit}
  \textfont\slfam=\tensl \def\sl{\fam\slfam\tensl}
  \textfont\ttfam=\tentt \scriptfont\ttfam=\eighttt
  \scriptscriptfont\ttfam=\eighttt  \def\tt{\fam\ttfam\tentt}
  \textfont\bffam=\tenbf \scriptfont\bffam=\sevenbf
  \scriptscriptfont\bffam=\fivebf \def\bf{\fam\bffam\tenbf}
     \ifx\arisposta\amsrisposta    \ifnum\contaeuler=1
  \textfont\eufmfam=\teneufm \scriptfont\eufmfam=\seveneufm
  \scriptscriptfont\eufmfam=\fiveeufm \def\eufm{\fam\eufmfam\teneufm}
  \textfont\eufbfam=\teneufb \scriptfont\eufbfam=\seveneufb
  \scriptscriptfont\eufbfam=\fiveeufb \def\eufb{\fam\eufbfam\teneufb}
  \def\eurm{\teneurm} \def\eurb{\teneurb} \def\eusm{\teneusm}
  \def\eusb{\teneusb}    \fi    \ifnum\contaams=1
  \textfont\msamfam=\tenmsam \scriptfont\msamfam=\sevenmsam
  \scriptscriptfont\msamfam=\fivemsam \def\msam{\fam\msamfam\tenmsam}
  \textfont\msbmfam=\tenmsbm \scriptfont\msbmfam=\sevenmsbm
  \scriptscriptfont\msbmfam=\fivemsbm \def\msbm{\fam\msbmfam\tenmsbm}
     \fi      \ifnum\contacyrill=1     \def\cyrill{\tenwncyr}
  \def\cyrilb{\tenwncyb}  \def\cyrili{\tenwncyi}         \fi
  \textfont3=\tenex \scriptfont3=\sevenex \scriptscriptfont3=\sevenex
  \def\cmmib{\fam\cmmibfam\tencmmib} \scriptfont\cmmibfam=\sevencmmib
  \textfont\cmmibfam=\tencmmib  \scriptscriptfont\cmmibfam=\fivecmmib
  \def\cmbsy{\fam\cmbsyfam\tencmbsy} \scriptfont\cmbsyfam=\sevencmbsy
  \textfont\cmbsyfam=\tencmbsy  \scriptscriptfont\cmbsyfam=\fivecmbsy
  \def\cmcsc{\fam\cmcscfam\tencmcsc} \scriptfont\cmcscfam=\eightcmcsc
  \textfont\cmcscfam=\tencmcsc \scriptscriptfont\cmcscfam=\eightcmcsc
     \fi            \tt \ttglue=.5em plus.25em minus.15em
  \normalbaselineskip=12pt
  \setbox\strutbox=\hbox{\vrule height8.5pt depth3.5pt width0pt}
  \let\sc=\eightrm \let\big=\tenbig   \normalbaselines
  \baselineskip=\infralinea  \rm}
\gdef\ninepoint{\def\rm{\fam0\ninerm}
  \textfont0=\ninerm \scriptfont0=\sixrm \scriptscriptfont0=\fiverm
  \textfont1=\ninei \scriptfont1=\sixi \scriptscriptfont1=\fivei
  \textfont2=\ninesy \scriptfont2=\sixsy \scriptscriptfont2=\fivesy
  \textfont3=\tenex \scriptfont3=\tenex \scriptscriptfont3=\tenex
  \def\mcal{\fam2 \ninesy}  \def\mmit{\fam1 \ninei}
  \textfont\itfam=\nineit \def\it{\fam\itfam\nineit}
  \textfont\slfam=\ninesl \def\sl{\fam\slfam\ninesl}
  \textfont\ttfam=\ninett \scriptfont\ttfam=\eighttt
  \scriptscriptfont\ttfam=\eighttt \def\tt{\fam\ttfam\ninett}
  \textfont\bffam=\ninebf \scriptfont\bffam=\sixbf
  \scriptscriptfont\bffam=\fivebf \def\bf{\fam\bffam\ninebf}
     \ifx\arisposta\amsrisposta  \ifnum\contaeuler=1
  \textfont\eufmfam=\nineeufm \scriptfont\eufmfam=\sixeufm
  \scriptscriptfont\eufmfam=\fiveeufm \def\eufm{\fam\eufmfam\nineeufm}
  \textfont\eufbfam=\nineeufb \scriptfont\eufbfam=\sixeufb
  \scriptscriptfont\eufbfam=\fiveeufb \def\eufb{\fam\eufbfam\nineeufb}
  \def\eurm{\nineeurm} \def\eurb{\nineeurb} \def\eusm{\nineeusm}
  \def\eusb{\nineeusb}     \fi   \ifnum\contaams=1
  \textfont\msamfam=\ninemsam \scriptfont\msamfam=\sixmsam
  \scriptscriptfont\msamfam=\fivemsam \def\msam{\fam\msamfam\ninemsam}
  \textfont\msbmfam=\ninemsbm \scriptfont\msbmfam=\sixmsbm
  \scriptscriptfont\msbmfam=\fivemsbm \def\msbm{\fam\msbmfam\ninemsbm}
     \fi       \ifnum\contacyrill=1     \def\cyrill{\ninewncyr}
  \def\cyrilb{\ninewncyb}  \def\cyrili{\ninewncyi}         \fi
  \textfont3=\nineex \scriptfont3=\sevenex \scriptscriptfont3=\sevenex
  \def\cmmib{\fam\cmmibfam\ninecmmib}  \textfont\cmmibfam=\ninecmmib
  \scriptfont\cmmibfam=\sixcmmib \scriptscriptfont\cmmibfam=\fivecmmib
  \def\cmbsy{\fam\cmbsyfam\ninecmbsy}  \textfont\cmbsyfam=\ninecmbsy
  \scriptfont\cmbsyfam=\sixcmbsy \scriptscriptfont\cmbsyfam=\fivecmbsy
  \def\cmcsc{\fam\cmcscfam\ninecmcsc} \scriptfont\cmcscfam=\eightcmcsc
  \textfont\cmcscfam=\ninecmcsc \scriptscriptfont\cmcscfam=\eightcmcsc
     \fi            \tt \ttglue=.5em plus.25em minus.15em
  \normalbaselineskip=11pt
  \setbox\strutbox=\hbox{\vrule height8pt depth3pt width0pt}
  \let\sc=\sevenrm \let\big=\ninebig \normalbaselines\rm}
\gdef\eightpoint{\def\rm{\fam0\eightrm}
  \textfont0=\eightrm \scriptfont0=\sixrm \scriptscriptfont0=\fiverm
  \textfont1=\eighti \scriptfont1=\sixi \scriptscriptfont1=\fivei
  \textfont2=\eightsy \scriptfont2=\sixsy \scriptscriptfont2=\fivesy
  \textfont3=\tenex \scriptfont3=\tenex \scriptscriptfont3=\tenex
  \def\mcal{\fam2 \eightsy}  \def\mmit{\fam1 \eighti}
  \textfont\itfam=\eightit \def\it{\fam\itfam\eightit}
  \textfont\slfam=\eightsl \def\sl{\fam\slfam\eightsl}
  \textfont\ttfam=\eighttt \scriptfont\ttfam=\eighttt
  \scriptscriptfont\ttfam=\eighttt \def\tt{\fam\ttfam\eighttt}
  \textfont\bffam=\eightbf \scriptfont\bffam=\sixbf
  \scriptscriptfont\bffam=\fivebf \def\bf{\fam\bffam\eightbf}
     \ifx\arisposta\amsrisposta   \ifnum\contaeuler=1
  \textfont\eufmfam=\eighteufm \scriptfont\eufmfam=\sixeufm
  \scriptscriptfont\eufmfam=\fiveeufm \def\eufm{\fam\eufmfam\eighteufm}
  \textfont\eufbfam=\eighteufb \scriptfont\eufbfam=\sixeufb
  \scriptscriptfont\eufbfam=\fiveeufb \def\eufb{\fam\eufbfam\eighteufb}
  \def\eurm{\eighteurm} \def\eurb{\eighteurb} \def\eusm{\eighteusm}
  \def\eusb{\eighteusb}       \fi    \ifnum\contaams=1
  \textfont\msamfam=\eightmsam \scriptfont\msamfam=\sixmsam
  \scriptscriptfont\msamfam=\fivemsam \def\msam{\fam\msamfam\eightmsam}
  \textfont\msbmfam=\eightmsbm \scriptfont\msbmfam=\sixmsbm
  \scriptscriptfont\msbmfam=\fivemsbm \def\msbm{\fam\msbmfam\eightmsbm}
     \fi       \ifnum\contacyrill=1     \def\cyrill{\eightwncyr}
  \def\cyrilb{\eightwncyb}  \def\cyrili{\eightwncyi}         \fi
  \textfont3=\eightex \scriptfont3=\sevenex \scriptscriptfont3=\sevenex
  \def\cmmib{\fam\cmmibfam\eightcmmib}  \textfont\cmmibfam=\eightcmmib
  \scriptfont\cmmibfam=\sixcmmib \scriptscriptfont\cmmibfam=\fivecmmib
  \def\cmbsy{\fam\cmbsyfam\eightcmbsy}  \textfont\cmbsyfam=\eightcmbsy
  \scriptfont\cmbsyfam=\sixcmbsy \scriptscriptfont\cmbsyfam=\fivecmbsy
  \def\cmcsc{\fam\cmcscfam\eightcmcsc} \scriptfont\cmcscfam=\eightcmcsc
  \textfont\cmcscfam=\eightcmcsc \scriptscriptfont\cmcscfam=\eightcmcsc
     \fi             \tt \ttglue=.5em plus.25em minus.15em
  \normalbaselineskip=9pt
  \setbox\strutbox=\hbox{\vrule height7pt depth2pt width0pt}
  \let\sc=\sixrm \let\big=\eightbig \normalbaselines\rm }
\gdef\tenbig#1{{\hbox{$\left#1\vbox to8.5pt{}\right.\n@space$}}}
\gdef\ninebig#1{{\hbox{$\textfont0=\tenrm\textfont2=\tensy
   \left#1\vbox to7.25pt{}\right.\n@space$}}}
\gdef\eightbig#1{{\hbox{$\textfont0=\ninerm\textfont2=\ninesy
   \left#1\vbox to6.5pt{}\right.\n@space$}}}
\def\alternativefont#1#2{\ifx\arisposta\amsrisposta \relax \else
\xdef#1{#2} \fi}
\global\contaeuler=0 \global\contacyrill=0 \global\contaams=0
%
%
%
%
\newbox\fotlinebb \newbox\hedlinebb \newbox\leftcolumn
\gdef\makeheadline{\vbox to 0pt{\vskip-22.5pt
     \fullline{\vbox to8.5pt{}\the\headline}\vss}\nointerlineskip}
\gdef\makehedlinebb{\vbox to 0pt{\vskip-22.5pt
     \fullline{\vbox to8.5pt{}\copy\hedlinebb\hfil
     \line{\hfill\the\headline\hfill}}\vss} \nointerlineskip}
\gdef\makefootline{\baselineskip=24pt \fullline{\the\footline}}
\gdef\makefotlinebb{\baselineskip=24pt
    \fullline{\copy\fotlinebb\hfil\line{\hfill\the\footline\hfill}}}
\gdef\doubleformat{\shipout\vbox{\Landspec\makehedlinebb
     \fullline{\box\leftcolumn\hfil\columnbox}\makefotlinebb}
     \advancepageno}
\gdef\columnbox{\leftline{\pagebody}}
\gdef\line#1{\hbox to\hsize{\hskip\leftskip#1\hskip\rightskip}}
\gdef\fullline#1{\hbox to\fullhsize{\hskip\leftskip{#1}%
\hskip\rightskip}}
\gdef\footnote#1{\let\@sf=\empty
	 \ifhmode\edef\#sf{\spacefactor=\the\spacefactor}\/\fi
	 #1\@sf\vfootnote{#1}}
\gdef\vfootnote#1{\insert\footins\bgroup
	 \ifnum\dimnota=1  \eightpoint\fi
	 \ifnum\dimnota=2  \ninepoint\fi
	 \ifnum\dimnota=0  \tenpoint\fi
	 \interlinepenalty=\interfootnotelinepenalty
	 \splittopskip=\ht\strutbox
	 \splitmaxdepth=\dp\strutbox \floatingpenalty=20000
	 \leftskip=\oldssposta \rightskip=\olddsposta
	 \spaceskip=0pt \xspaceskip=0pt
	 \ifnum\sinnota=0   \textindent{#1}\fi
	 \ifnum\sinnota=1   \item{#1}\fi
	 \footstrut\futurelet\next\fo@t}
\gdef\fo@t{\ifcat\bgroup\noexpand\next \let\next\f@@t
	     \else\let\next\f@t\fi \next}
\gdef\f@@t{\bgroup\aftergroup\@foot\let\next}
\gdef\f@t#1{#1\@foot} \gdef\@foot{\strut\egroup}
\gdef\footstrut{\vbox to\splittopskip{}}
\skip\footins=\bigskipamount
\count\footins=1000  \dimen\footins=8in
\catcode`@=12
\tenpoint
\ifnum\unoduecol=1 \hsize=\tothsize   \fullhsize=\tothsize \fi
\ifnum\unoduecol=2 \hsize=\collhsize  \fullhsize=\tothsize \fi
\global\let\lrcol=L      \ifnum\unoduecol=1
\output{\plainoutput{\ifnum\tipbnota=2 \clearnmbnota\fi}} \fi
\ifnum\unoduecol=2 \output{\if L\lrcol
     \global\setbox\leftcolumn=\columnbox
     \global\setbox\fotlinebb=\line{\hfill\the\footline\hfill}
     \global\setbox\hedlinebb=\line{\hfill\the\headline\hfill}
     \advancepageno  \global\let\lrcol=R
     \else  \doubleformat \global\let\lrcol=L \fi
     \ifnum\outputpenalty>-20000 \else\dosupereject\fi
     \ifnum\tipbnota=2\clearnmbnota\fi }\fi
\def\ifdoublepage{\ifnum\unoduecol=2 }
\gdef\yespagenumbers{\footline={\hss\tenrm\folio\hss}}
\gdef\ciao{ \ifnum\fdefcontre=1 \endfdef\fi
     \par\vfill\supereject \ifnum\unoduecol=2
     \if R\lrcol  \headline={}\nopagenumbers\null\vfill\eject
     \fi\fi \end}

\newskip\olddsposta \newskip\oldssposta
\global\oldssposta=\leftskip \global\olddsposta=\rightskip

\def\filldots{\leaders\hbox to 1em{\hss.\hss}\hfill}
\def\inquadrb#1 {\vbox {\hrule  \hbox{\vrule \vbox {\vskip .2cm
    \hbox {\ #1\ } \vskip .2cm } \vrule  }  \hrule} }
 \def\newline{\hfil\break}
\def\jump{\vskip\baselineskip} \newskip\iinnffrr
\def\sjump{\iinnffrr=\baselineskip
	  \divide\iinnffrr by 2 \vskip\iinnffrr}
\def\bjump{\vskip\baselineskip \vskip\baselineskip}
\newcount\nmbnota  \def\clearnmbnota{\global\nmbnota=0}
\newcount\tipbnota \def\letterfootnote{\global\tipbnota=1}

\def\note#1{\global\advance\nmbnota by 1 \ifnum\tipbnota=1
    \footnote{$^{\rm\nttlett}$}{#1} \else {\ifnum\tipbnota=2
    \footnote{$^{\nttsymb}$}{#1}
    \else\footnote{$^{\the\nmbnota}$}{#1}\fi}\fi}
\def\nttlett{\ifcase\nmbnota \or a\or b\or c\or d\or e\or f\or
g\or h\or i\or j\or k\or l\or m\or n\or o\or p\or q\or r\or
s\or t\or u\or v\or w\or y\or x\or z\fi}
\def\nttsymb{\ifcase\nmbnota \or\dag\or\sharp\or\ddag\or\star\or
\natural\or\flat\or\clubsuit\or\diamondsuit\or\heartsuit
\or\spadesuit\fi}   \clearnmbnota
\def\numberfootnote{\global\tipbnota=0} \numberfootnote
\def\setnote#1{\expandafter\xdef\csname#1\endcsname{
\ifnum\tipbnota=1 {\rm\nttlett} \else {\ifnum\tipbnota=2
{\nttsymb} \else \the\nmbnota\fi}\fi} }
\newcount\nbmfig  \def\clearnbmfig{\global\nbmfig=0}
\gdef\figure{\global\advance\nbmfig by 1
      {\rm fig. \the\nbmfig}}   \clearnbmfig
\def\setfig#1{\expandafter\xdef\csname#1\endcsname{fig. \the\nbmfig}}
 \def\endformula{\eqno\numero $$}
 \def\efr{\endformula}
\newcount\frmcount \def\clearfrmcount{\global\frmcount=0}
\def\numero{\global\advance\frmcount by 1   \ifnum\indappcount=0
  {\ifnum\cpcount <1 {\hbox{\rm (\the\frmcount )}}  \else
  {\hbox{\rm (\the\cpcount .\the\frmcount )}} \fi}  \else
  {\hbox{\rm (\applett .\the\frmcount )}} \fi}
\def\nfr{\nameformula}    
\def\nameformula#1{\global\advance\frmcount by 1%
{\ifnum\indappcount=0%
{\ifnum\cpcount<1\xdef\spzzttrra{(\the\frmcount )}%
\else\xdef\spzzttrra{(\the\cpcount .\the\frmcount )}\fi}%
\else\xdef\spzzttrra{(\applett .\the\frmcount )}\fi}%
\expandafter\xdef\csname#1\endcsname{\spzzttrra}%
\eqno{\ifnum\draftnum=0\hbox{\rm\spzzttrra}\else%
\hbox{$\buildchar{\rm\spzzttrra}{\tt\scriptscriptstyle#1}{}$}\fi}$$}
\def\nameali#1{\global\advance\frmcount by 1%
{\ifnum\indappcount=0%
{\ifnum\cpcount<1\xdef\spzzttrra{(\the\frmcount )}%
\else\xdef\spzzttrra{(\the\cpcount .\the\frmcount )}\fi}%
\else\xdef\spzzttrra{(\applett .\the\frmcount )}\fi}%
\expandafter\xdef\csname#1\endcsname{\spzzttrra}%
\ifnum\draftnum=0\hbox{\rm\spzzttrra}\else%
\hbox{$\buildchar{\rm\spzzttrra}{\tt\scriptscriptstyle#1}{}$}\fi}
\clearfrmcount
\newcount\cpcount \def\clearcpcount{\global\cpcount=0}
\newcount\subcpcount \def\clearsubcpcount{\global\subcpcount=0}
\newcount\appcount \def\clearappcount{\global\appcount=0}
\newcount\indappcount \def\clearindappcount{\indappcount=0}
\newcount\sottoparcount 

\def\applett{\ifcase\appcount  \or {A}\or {B}\or {C}\or
{D}\or {E}\or {F}\or {G}\or {H}\or {I}\or {J}\or {K}\or {L}\or
{M}\or {N}\or {O}\or {P}\or {Q}\or {R}\or {S}\or {T}\or {U}\or
{V}\or {W}\or {X}\or {Y}\or {Z}\fi    \ifnum\appcount<0
\immediate\write16 {Panda ERROR - Appendix: counter "appcount"
out of range}\fi  \ifnum\appcount>26  \immediate\write16 {Panda
ERROR - Appendix: counter "appcount" out of range}\fi}
\clearappcount  \clearindappcount \newcount\connttrre
\def\clearconnttrre{\global\connttrre=0} \newcount\countref
\def\clearcountref{\global\countref=0} \clearcountref
\def\chapter#1{\global\advance\cpcount by 1 \clearfrmcount
		 \goodbreak\null\vbox{\jump\nobreak
		 \clearsubcpcount\clearindappcount
		 \itemitem{\ttaarr\the\cpcount .\qquad}{\ttaarr #1}
		 \par\nobreak\jump\sjump}\nobreak}
\def\section#1{\global\advance\subcpcount by 1 \goodbreak\null
	       \vbox{\sjump\nobreak\ifnum\indappcount=0
		 {\ifnum\cpcount=0 {\itemitem{\ppaarr
	       .\the\subcpcount\quad\enskip\ }{\ppaarr #1}\par} \else
		 {\itemitem{\ppaarr\the\cpcount .\the\subcpcount\quad
		  \enskip\ }{\ppaarr #1} \par}  \fi}
		\else{\itemitem{\ppaarr\applett .\the\subcpcount\quad
		 \enskip\ }{\ppaarr #1}\par}\fi\nobreak\jump}\nobreak}
\clearsubcpcount
\def\appendix#1{\global\advance\appcount by 1 \clearfrmcount
		  \goodbreak\null\vbox{\jump\nobreak
		  \global\advance\indappcount by 1 \clearsubcpcount
	  \itemitem{ }{\hskip-40pt\ttaarr Appendix\ \applett :\ #1}
	     \nobreak\jump\sjump}\nobreak}
\clearappcount \clearindappcount
\def\references{\goodbreak\null\vbox{\jump\nobreak
   \itemitem{}{\ttaarr References} \nobreak\jump\sjump}\nobreak}

\clearcpcount\clearcountref

\def\setchap#1{\ifnum\indappcount=0{\ifnum\subcpcount=0%
\xdef\spzzttrra{\the\cpcount}%
\else\xdef\spzzttrra{\the\cpcount .\the\subcpcount}\fi}
\else{\ifnum\subcpcount=0 \xdef\spzzttrra{\applett}%
\else\xdef\spzzttrra{\applett .\the\subcpcount}\fi}\fi
\expandafter\xdef\csname#1\endcsname{\spzzttrra}}
\newcount\draftnum \newcount\ppora   \newcount\ppminuti
\global\ppora=\time   \global\ppminuti=\time
\global\divide\ppora by 60  \draftnum=\ppora
\multiply\draftnum by 60    \global\advance\ppminuti by -\draftnum
\def\droggi{\number\day /\number\month /\number\year\ \the\ppora
:\the\ppminuti}     \global\draftnum=0
\def\draftcomment#1{\ifnum\draftnum=0 \relax \else
{\ {\bf ***}\ #1\ {\bf ***}\ }\fi} 
%
%
\catcode`@=11
\gdef\Ref#1{\expandafter\ifx\csname @rrxx@#1\endcsname\relax%
{\global\advance\countref by 1    \ifnum\countref>200
\immediate\write16 {Panda ERROR - Ref: maximum number of references
exceeded}  \expandafter\xdef\csname @rrxx@#1\endcsname{0}\else
\expandafter\xdef\csname @rrxx@#1\endcsname{\the\countref}\fi}\fi
\ifnum\draftnum=0 \csname @rrxx@#1\endcsname \else#1\fi}
\gdef\beginref{\ifnum\draftnum=0  \gdef\Rref{\fairef}
\gdef\endref{\scriviref} \else\relax\fi
\ifx\risposta\mplarisposta \ninepoint \fi
\baselineskip=12pt \parskip 2pt plus.2pt }
\def\Reflab#1{[#1]} \gdef\Rref#1#2{\item{\Reflab{#1}}{#2}}
\gdef\endref{\relax}  \newcount\conttemp
\gdef\fairef#1#2{\expandafter\ifx\csname @rrxx@#1\endcsname\relax
{\global\conttemp=0 \immediate\write16 {Panda ERROR - Ref: reference
[#1] undefined}} \else
{\global\conttemp=\csname @rrxx@#1\endcsname } \fi
\global\advance\conttemp by 50  \global\setbox\conttemp=\hbox{#2} }
\gdef\scriviref{\clearconnttrre\conttemp=50
\loop\ifnum\connttrre<\countref \advance\conttemp by 1
\advance\connttrre by 1
\item{\Reflab{\the\connttrre}}{\unhcopy\conttemp} \repeat}
\clearcountref \clearconnttrre
\catcode`@=12
\ifx\risposta\mplarisposta \def\Reflab#1{#1.} \letterfootnote \fi
%
%

\def\slashchar#1{\setbox0=\hbox{$#1$} \dimen0=\wd0
     \setbox1=\hbox{/} \dimen1=\wd1 \ifdim\dimen0>\dimen1
      \rlap{\hbox to \dimen0{\hfil/\hfil}} #1 \else
      \rlap{\hbox to \dimen1{\hfil$#1$\hfil}} / \fi}
\ifx\oldchi\undefined \let\oldchi=\chi
  \def\cchi{{\raise 1pt\hbox{$\oldchi$}}} \let\chi=\cchi \fi
\ifnum\contasym=1 \else \fi
 \def\del{\partial}   

\def\frac#1#2{{\textstyle{#1 \over #2}}}

\def\half{\ifinner {\scriptstyle {1 \over 2}}\else {1 \over 2} \fi}

\def\simge{\rlap{\raise 2pt \hbox{$>$}}{\lower 2pt \hbox{$\sim$}}}
\def\simle{\rlap{\raise 2pt \hbox{$<$}}{\lower 2pt \hbox{$\sim$}}}

\def\buildchar#1#2#3{{\null\!\mathop{#1}\limits^{#2}_{#3}\!\null}}

\def\vbig#1#2{{\vbigd@men=#2\divide\vbigd@men by 2%
\hbox{$\left#1\vbox to \vbigd@men{}\right.\n@space$}}}

\def\noblackbox{\overfullrule=0pt} 
%
%
\newcount\fdefcontre \newcount\fdefcount \newcount\indcount
\newread\filefdef  \newread\fileftmp  \newwrite\filefdef
\newwrite\fileftmp     \def\strip #1*.A {#1}%
\def\futuredef#1{\beginfdef
\expandafter\ifx\csname#1\endcsname\relax%
{\immediate\write\fileftmp{#1*.A}%
\immediate\write16 {Panda Warning - fdef: macro "#1" on page
\the\pageno \space undefined}
\ifnum\draftnum=0 \expandafter\xdef\csname#1\endcsname{(?)}
\else \expandafter\xdef\csname#1\endcsname{(#1)}\fi
\global\advance\fdefcount by 1}\fi\csname#1\endcsname}

\def\beginfdef{\ifnum\fdefcontre=0
\immediate\openin\filefdef\jobname.fdef
\immediate\openout\fileftmp\jobname.ftmp
\global\fdefcontre=1  \ifeof\filefdef \immediate\write16 {Panda
WARNING - fdef: file \jobname.fdef not found, run TeX again}
\else \immediate\read\filefdef to\spzzttrra
\global\advance\fdefcount by \spzzttrra
\indcount=0 \loop\ifnum\indcount<\fdefcount
\advance\indcount by 1%
\immediate\read\filefdef to\spezttrra%
\immediate\read\filefdef to\sppzttrra%
\edef\spzzttrra{\expandafter\strip\spezttrra}%
\immediate\write\fileftmp {\spzzttrra *.A}
\expandafter\xdef\csname\spzzttrra\endcsname{\sppzttrra}%
\repeat \fi \immediate\closein\filefdef \fi}
\def\endfdef{\immediate\closeout\fileftmp   \ifnum\fdefcount>0
\immediate\openin\fileftmp \jobname.ftmp
\immediate\openout\filefdef \jobname.fdef
\immediate\write\filefdef {\the\fdefcount}   \indcount=0
\loop\ifnum\indcount<\fdefcount    \advance\indcount by 1
\immediate\read\fileftmp to\spezttrra
\edef\spzzttrra{\expandafter\strip\spezttrra}
\immediate\write\filefdef{\spzzttrra *.A}
\edef\spezttrra{\string{\csname\spzzttrra\endcsname\string}}
\iwritel\filefdef{\spezttrra}
\repeat  \immediate\closein\fileftmp \immediate\closeout\filefdef
\immediate\write16 {Panda Warning - fdef: Label(s) may have changed,
re-run TeX to get them right}\fi}
\def\iwritel#1#2{\newlinechar=-1
{\newlinechar=`\ \immediate\write#1{#2}}\newlinechar=-1}
\global\fdefcontre=0 \global\fdefcount=0 \global\indcount=0
%
%
%
\mathchardef\alpha="710B   \mathchardef\beta="710C
\mathchardef\gamma="710D   \mathchardef\delta="710E
\mathchardef\epsilon="710F   \mathchardef\zeta="7110
\mathchardef\eta="7111   \mathchardef\theta="7112
\mathchardef\iota="7113   \mathchardef\kappa="7114
\mathchardef\lambda="7115   \mathchardef\mu="7116
\mathchardef\nu="7117   \mathchardef\xi="7118
\mathchardef\pi="7119   \mathchardef\rho="711A
\mathchardef\sigma="711B   \mathchardef\tau="711C
\mathchardef\upsilon="711D   \mathchardef\phi="711E
\mathchardef\chi="711F   \mathchardef\psi="7120
\mathchardef\omega="7121   \mathchardef\varepsilon="7122
\mathchardef\vartheta="7123   \mathchardef\varpi="7124
\mathchardef\varrho="7125   \mathchardef\varsigma="7126
\mathchardef\varphi="7127
%
%
\null
%
%
%
%
%
\loadamsmath
\chapterfont{\bfone} \sectionfont{\scaps}
\noblackbox
\def\eqmodtwo{\ \buildchar{=}{{\rm{\scriptscriptstyle MOD\ 2}}}{ }\ }
\nopagenumbers
{\baselineskip=12pt
\line{\hfill NBI-HE-95-06}
\line{\hfill hep-th/9503040}
\line{\hfill March, 1995}}
{\baselineskip=14pt
\vfill
\centerline{\capsone Bosonization of World-Sheet Fermions}
\sjump
\centerline{\capsone in Minkowski Space-Time.}
\bjump\bjump
\centerline{\scaps Andrea Pasquinucci~\footnote{$^\dagger$}{Supported
by EU grant no. ERBCHBGCT920179.} and Kaj
Roland~\footnote{$^\ddagger$}{Supported by the Carlsberg Foundation.}}
\sjump
\centerline{\sl The Niels Bohr Institute, University of Copenhagen,}
\centerline{\sl Blegdamsvej 17, DK-2100, Copenhagen, Denmark}
\bjump \vfill
\centerline{\capsone ABSTRACT}
\sjump
\noindent
We propose a way of bosonizing free world-sheet fermions for
$4$-dimensional heterotic string theory formulated in Minkowski
space-time. We discuss the differences as compared to the standard
bosonization performed in Euclidean space-time.
\sjump \vfill
\pageno=0 \eject }
\yespagenumbers\pageno=1
\null\bjump
{\bf 1.} In the Neveu-Schwarz-Ramond approach to superstring
theory~[\Ref{GSW}]
any model with $D$ flat space-time dimensions will contain the
coordinate fields $X^{\mu}$ and their
world-sheet superpartners, the set of Majorana fermion
fields $\psi^{\mu}$. The latter can have either Neveu-Schwarz (NS) or
Ramond (R) boundary conditions around any non-contractible loop on the
world-sheet. In order to obtain string scattering amplitudes one has
to compute correlation functions of these fermions on arbitrary Riemann
surfaces. The standard way of doing this is by bosonization.
Bosonization is usually carried out in Euclidean space-time, and
thus requires that we first rotate the metric from Minkowski to
Euclidean space-time, then perform all calculations and only at the end
rotate back to Minkowski space-time.

On the other hand, it could be convenient to formulate the theory
and make all computations directly in Minkowski space-time.
For example, the whole concept of unitarity requires that we are in
Minkowski space-time, even though, to obtain amplitudes with the correct
analytic properties, a careful procedure of analytic continuation in the
momentum invariants is needed, as discussed for example in ref.~[\Ref{DP}].
Likewise, an important quantity such as the
time-reversal operator $T$ only assumes its true physical significance
in Minkowski space-time. Accordingly, if we want to prove the CPT-invariance
of the $S$-matrix in a given string theory at any order in perturbation
theory, as was done in ref.~[\Ref{Sonoda}] for the 10-dimensional heterotic
models, it would be most natural to define the CPT operator and
make the proof entirely in  Minkowski space-time.
So, in some situations it might be useful if the bosonization could be
carried out directly in Minkowski space-time, without having to rotate
the metric.

In this letter we propose a bosonization procedure in Minkowski space-time.
In section 2 we consider the simplest possible case, corresponding to
a single pair of Majorana fermions transforming in the vector
representation of $SO(1,1)$, and in sections 3 and 4 we then
discuss how to incorporate other fermions, including the proper treatment of
cocycles. To be explicit we consider the $4$-dimensional heterotic
string models of Kawai, Lewellen and Tye (KLT)~[\Ref{KLT}], but our
procedure should apply to other models as well.

\sjump

{\bf 2.} We start by reviewing very briefly the well-known case of
bosonization in $D=2$ Euclidean space-time, that is,
we consider two free chiral Majorana fermions
on the world-sheet, transforming as vectors under $SO(2)$. In terms of a
local complex coordinate $z$ they are represented by two hermitean
chiral conformal fields $\psi^{\mu}(z)$ of dimension $1/2$,
with operator product expansion (OPE)
$$ \psi^{\mu} (z) \psi^{\nu} (w) = g^{\mu \nu} {1 \over z-w} + \ldots
\qquad\qquad \mu,\nu=(0,1) \ ,
\nfr{EOPE}
with $g^{\mu\nu}=\delta^{\mu\nu}$.

(In general the hermitean conjugate field
$\phi_{\Delta}^{*}$ of a primary conformal field $\phi_{\Delta}(z)$ of
dimension $\Delta$ is defined by the relation
$$ (\phi_{\Delta} (z) )^{\dagger} = \left( {1 \over z^{*}}
\right)^{2\Delta} \phi^{*}_{\Delta} \left(
{1 \over z^{*}} \right) \ , \nfr{herm}
where $z^{*}$ denotes the complex conjugate of $z$, and we say that
$\phi_{\Delta}$ is hermitean (anti-hermitean) if $\phi_{\Delta}^{*} =
\phi_{\Delta}$ ($\phi_{\Delta}^{*} =
-\phi_{\Delta}$).)

In Euclidean space-time
we bosonize by introducing an anti-hermitean
scalar field $\phi$, defined by $j^{01}(z) \equiv -i \psi^0 \psi^1 (z) =
\partial \phi (z)$, whose mode operators give rise to a Fock space of states
with positive definite norm. We may then identify $\psi^{\pm} = e^{\pm
\phi}$ where $\psi^{\pm}$ are two {\it complex \/}
fermions formed out of $\psi^0$ and $\psi^1$.~\note{Products
(and exponentials) of operators at the same
point are always normal ordered. We do not adopt the $: \ \  :$ notation.}
The situation is summarized in Table 1.
{\topinsert
\setbox\strutbox=\hbox{\vrule height16pt depth3.5pt width0pt} 
$$\vbox{\offinterlineskip \halign{\strut#& \vrule# &\hfill #\hfill &
\vrule# & \hfill #\hfill & \vrule# \cr
\noalign{\hrule}
&\ &\ Euclidean\ &\ &\ Minkowski\ &\cr
\noalign{\hrule}
&& $\psi^{0} = {1 \over {\bf i}\sqrt{2}} \left( e^{\phi} -
   e^{-\phi} \right)$ && $\psi^{0} = {1 \over \sqrt{2}}
   \left( e^{\phi} - e^{-\phi} \right)$ &\cr
&& $\psi^{1} = {1 \over \sqrt{2}} \left( e^{\phi} +
   e^{-\phi} \right)$ && $\psi^{1} = {1\over\sqrt{2}} \left( e^{\phi} +
   e^{-\phi} \right)$ &\cr
&& $e^{\pm\phi} = \psi^\pm = {1\over \sqrt{2}} (\psi^1\pm{\bf i}\psi^0)$
   &&$e^{\pm\phi} = \psi^\pm = {1\over \sqrt{2}} (\psi^1\pm \psi^0)$ &\cr
&& $\psi^- =(\psi^+)^*$ &&  $(\psi^{\pm})^* = \psi^{\pm}$ &\cr
&& $j^{01}(z)=\del\phi(z)$ && $j^{01}(z)={\bf -i}\del\phi(z)$ &\cr
\noalign{\hrule}
&& \multispan3 \hfill $\phi(z)\phi(w) = +\log(z-w) + \ldots$\hfill &\cr
&& \multispan3 \hfill
   $\phi(z) = x + N \log z + \sum_{n \neq 0} {\alpha_n \over n} z^{-n}$
   \hfill &\cr
&& \multispan3 \hfill
   $[\alpha_n, \alpha_m] = n \delta_{n+m}$ \ , \hfill $[N,x] = 1 $
   \hfill &\cr
\noalign{\hrule}
&& $ \phi $\ \ {\bf anti-hermitean} && $\phi$\ \ {\bf hermitean}  &\cr
&& $(\alpha_n)^{\dagger} = {\bf +}\alpha_{-n}$ &&
   $(\alpha_n)^{\dagger} = {\bf -} \alpha_{-n}$ &\cr
&& $x^{\dagger} = {\bfmath -} x$ &&$x^{\dagger} = {\bf +} x$ &\cr
&& $N^{\dagger} = {\bfmath +} N $ &&$N^{\dagger} = {\bf -} N $ &\cr
&& $\langle N = r \vert N = r' \rangle ={\bfmath \delta_{\bf
   r{\bfmath-}r'}} $ && $\langle N=r\vert N = r'\rangle =
   {\bfmath\delta_{\bf r\bfmath{+}r'}} $ &\cr
\noalign{\hrule}
}}$$
\centerline{{\bf Table 1:} Comparison of the bosonization in
Euclidean/Minkowski space-time}
\centerline{for two free chiral Majorana fermions.}
\sjump
\endinsert}
Notice that in this case the ``momentum'' operator
$N$ is hermitean. In particular, if we define
$ \vert N = r \rangle = \lim_{z \rightarrow 0} e^{r\phi(z)} \vert 0
\rangle $,
we find by conservation of the ``momentum''
$$\eqalignno{
& \langle N = r \vert N = r' \rangle \ = \ \lim_{\zeta \rightarrow 0}
\lim_{z \rightarrow 0} \langle \left( e^{r \phi(\zeta)}
\right)^{\dagger} e^{r' \phi(z)} \rangle & \nameali{innerproduct} \cr
&\qquad \ = \lim_{\zeta \rightarrow 0}\lim_{z \rightarrow 0}
\langle \left( \zeta^{*} \right)^{-r^2/2} e^{-r\phi(1/\zeta^{*})}
\ e^{r' \phi(z)} \rangle \ = \ \delta_{r-r'} \qquad ({\rm Euclidean \ case})
\ . \cr } $$

In Minkowski space-time, the metric $g$ appearing in the OPE \EOPE\
becomes
$g={\rm diag}_{(2)}(-1,1)$. To bosonize $\psi^\mu$
in this case we note that the Ka\v{c}-Moody current $j^{01} = -i \psi^0
\psi^1$ remains hermitean
but now has the opposite sign in the OPE compared to the Euclidean case.
This means that the scalar field
$\phi$, just like the time
coordinate field $X^{0}$, is forced to have the ``wrong'' sign in the
OPE and thus gives rise to a Fock space containing states of negative norm.
We choose to define $j^{01}(z) \equiv -i \partial \phi (z)$ so that
$\phi$ is now hermitean (rather than anti-hermitean)~\note{It is also
possible, although not
convenient, to choose $\phi$ anti-hermitean, again with the ``wrong''
sign in the OPE, $\phi(z) \phi(w) = -\log (z-w) + \ldots $~.}
and then the OPE with the ``wrong'' sign is
$$ \phi(z) \phi(w) = +\log (z-w) + \ldots  \ .
\nfr{opeboson}
The bosonization proceeds as before (see Table 1), but the
hermiticity properties of the operators are different. The operators
$e^{\pm \phi} \equiv \psi^{\pm}$ are now hermitean and have to be
identified with {\it hermitean} linear combinations of $\psi^0$ and
$\psi^1$. Also,
the ``momentum'' operator $N$ is {\it anti-hermitean\/}.
At first sight this seems to lead to an inconsistency: Generically an
anti-hermitean operator should have imaginary eigenvalues. Instead,
the states in the Neveu-Schwarz (NS) sector (i.e. those that
are created from the conformal vacuum by operators obtained from the fields
$\psi^{\mu}$ and their derivatives) involve only real integer values of
the ``momentum'' $N$. Similarly, in the Ramond (R) sector,
the eigenvalues are half-integer real numbers.
An anti-hermitean operator can
have nonzero real eigenvalues only if the corresponding
eigenstates have zero norm. But
this is exactly the case! If we redo the computation leading to
equation \innerproduct\ then, since $\phi$ is now hermitean rather than
anti-hermitean, we obtain
\hbox{$\langle N=r\vert N = r'\rangle ={\bfmath\delta_{\bf r\bfmath{+}r'}}$}
for any real numbers $r,r'$. In particular
\hbox{$\langle N=r \vert N=r \rangle = 0$} for any real $r\neq 0$.

As is well known there are two ground states in the Ramond sector,
$\vert \pm 1/2 \rangle$, created from the conformal vacuum by the spin
field operators
$$ S_{\pm 1/2} (z) \equiv e^{\pm \phi(z)/2} \ , \nfr{spinfield}
which transform in the spinor representation of $SO(1,1)$.
The corresponding Clifford algebra is generated by the matrices
$\gamma^{\mu}$ defined by the OPE
$$ \psi^{\mu} (z) S_{\alpha} (w) = {1 \over \sqrt{2}} {1 \over
\sqrt{z-w}} (\gamma^{\mu})_{\alpha}^{\ \beta} S_{\beta} (w) + \ldots \ .
\nfr{cliffordOPE}
Explicitly one finds $ \gamma^0 = -i \sigma_2$ and $\gamma^1 = \sigma_1$.

This concludes our discussion of the bosonization for a single pair of
chiral $SO(1,1)$ Majorana fermions. In passing we note that all
correlation functions on an arbitrary Riemann
surface are the same whether $\phi$ is assumed to be
hermitean (as in the Minkowski case) or anti-hermitean
(as in the Euclidean case). This is because the
correlation functions are uniquely determined by the boundary conditions
(i.e. the spin structures) together with the short distance behaviour
(i.e. the OPE) which are the same in both cases. Thus, the $N$-point $g$-loop
vertex of ref.~[\Ref{PDV1}] can be used also in the Minkowski case.

\sjump

{\bf 3.} When there are other fermions present besides $\psi^{0}$ and
$\psi^{1}$, it is
necessary to introduce cocycles in order to ensure that different
fermions anti-commute even after they have been bosonized~[\Ref{Koste}].
Furthermore,
the typical spin field neither commutes nor anti-commutes with other
operators but picks up a phase that is a fractional power of $-1$, and
the cocycles serve to keep track also of this.

A priori there are many ways of choosing the cocycle operators, but the
choices are limited by a number of consistency conditions. For the
sake of being definite we will consider four-dimensional heterotic string
models of the type described by Kawai, Lewellen and Tye [\Ref{KLT}],
but the generalization to other models is straightforward. In ref.
[\Ref{ammedm}] we discussed in great detail how to bosonize a generic
4-dimensional heterotic KLT model in Euclidean space-time and here we
briefly summarize the main points.

In a 4-dimensional heterotic KLT model in Euclidean space-time,
there are 22 left-moving complex fermions
$\bar{\psi}_{(\bar{l})}^+ (\bar{z})$, $\bar{l}=\bar{1}, \ldots,
\overline{22}$, and eleven right-moving complex fermions $\psi_{(l)}^+$,
$l=0,1,\ldots,10$, with $(\bar{\psi}_{(\bar{l})}^+)^* =
\bar{\psi}_{(\bar{l})}^-$
and $(\psi^+_{(l)})^* = \psi^-_{(l)}$. As usual, a set of hermitean
fermions $\psi_{(11)}^{\pm}$ is introduced when we fermionize the
superghosts in the usual way, $\beta = \partial \xi \, \psi_{(11)}^-$ and
$\gamma = \psi^+_{(11)} \eta$.

{\topinsert
\setbox\strutbox=\hbox{\vrule height16pt depth3.5pt width0pt} 
$$\vbox{\offinterlineskip \halign{\strut#& \vrule# &\hfill #\hfill &
\vrule# & \hfill #\hfill & \vrule# \cr
\noalign{\hrule}
&\ &\ Euclidean\ &\ &\ Minkowski\ &\cr
\noalign{\hrule}
&& $\psi^{\mu} (z) \psi^{\nu} (w) = {\displaystyle\delta^{\mu \nu}
   \over\displaystyle z-w}$,  $\mu,\nu=(0,\ldots,3)$ &&
   $\psi^{\mu} (z) \psi^{\nu} (w) = {\displaystyle g^{\mu \nu}
   \over\displaystyle z-w}$, ~ $g=(-,+,+,+)$ &\cr
&& $\psi^\mu={1\over{\sqrt{2}}}\Big\{ {1\over{\bf i}}(\psi^+_{(0)} -
   \psi^-_{(0)}), (\psi^+_{(0)} +\psi^-_{(0)}),$ &&
   $\psi^\mu={1\over{\sqrt{2}}}\Big\{ (\psi^+_{(0)} -
   \psi^-_{(0)}), (\psi^+_{(0)} +\psi^-_{(0)}),$ &\cr
&& $~~~~~{1\over{\bf i}}(\psi^+_{(1)}-\psi^-_{(1)}),
   (\psi^+_{(1)} +\psi^-_{(1)})\Big\}$ &&
   $~~~~~(\psi^+_{(1)} -\psi^-_{(1)}), (\psi^+_{(1)}+
   \psi^-_{(1)})\Big\}$ &\cr
&& $\Psi_{(L)}^- = (\Psi_{(L)}^+)^* \ \ L=1,\ldots,{\bf 33}$
   &&$\Psi_{(L)}^- = (\Psi_{(L)}^+)^*\ \ L=1,\ldots,{\bf 32}$
   &\cr
&& $\Psi_{(L)}^\pm = (\Psi_{(L)}^\pm)^* \ \ L={\bf 34}$ &&
   $\Psi_{(L)}^\pm = (\Psi_{(L)}^\pm)^* \ \ L={\bf 33,34}$ &\cr
\noalign{\hrule}
&& \multispan3 \hfill
   $\Psi_{(L)}^\pm = e^{\pm\Phi_{(L)}} \left(C_{(L)}\right)^{\pm 1}$
   \hfill&\cr
&& \multispan3 \hfill
   $\Phi_{(L)} (z) \Phi_{(K)} (w) = \eta_{L,K} \log (z-w) + \ldots$
   \ ,\hfill
   $\eta_{L,K} = {\rm diag}_{(34)}(+,\ldots,+,-)$
   \hfill &\cr
&& \multispan3 \hfill
   $ [ J_0^{(L)}, \Phi_{(K)} ] = \delta_{L,K}$\ , \hfill
   $( J_0^{(34)})^{\dagger} = - J_0^{(34)} -2 $
   \hfill &\cr
\noalign{\hrule}
&& $C_{(L)}=C_{\rm gh}^{(L)}\cdot e^{i \pi e_{(L)}\cdot Y^E\cdot J_0 }$
  &&$C_{(L)}=C_{\rm gh}^{(L)}\cdot e^{i\pi e_{(L)}\cdot Y^M\cdot J_0 }$ &\cr
&& $(J_0^{(L)})^{\dagger}=J_0^{(L)}\qquad L=1,\ldots,{\bf 33}$
   &&$(J_0^{(L)})^{\dagger}=J_0^{(L)}\qquad L=1,\ldots,{\bf 32}$
   &\cr
&& ------ &&
   $( J_0^{(33)})^{\dagger} = {\bfmath -} J_0^{(33)} $ &\cr
&& $(C_{(L)})^{\dagger} = (C_{(L)})^{-1}\qquad L=1,\ldots,{\bf 33} $ &&
   $(C_{(L)})^{\dagger} = (C_{(L)})^{-1}\qquad L=1,\ldots,{\bf 32} $ &\cr
\noalign{\hrule}
}}$$
\centerline{{\bf Table 2:}  Comparison of the bosonization in
Euclidean/Minkowski space-time}
\centerline{for a 4d KLT heterotic string model.}
\sjump
\endinsert}

To define the cocycles we have to introduce an {\it ordering} for the 33
complex fermions and $\psi_{(11)}^+$, i.e. number them by integers
$L$ running from $1$ to $34$. The most natural example of an ordering is
$( \bar{1}, \bar{2} , \ldots ,\overline{22}; 0, 1, \ldots
,10;11 )$ where the left-moving
fermions $\bar{\psi}_{(\bar{l})}^\pm$ are represented by the number $L=l$,
while the right-moving fermions $\psi_{(l)}^\pm$ are represented by the
number $L=22+l$ and the superghost fermions $\psi_{(11)}^{\pm}$ by $L=34$.

Given such an ordering it is then convenient to denote the
fermions corresponding to the integer $L$ by $\Psi_{(L)}^\pm$.

The most relevant bosonization formulae are summarized in Table 2, and
more details can be found in ref.~[\Ref{ammedm}]. In the
expression for the cocycle $C_{(L)}$ the
$34$ dimensional vector $e_{(L)}$ is given by $e_{(L)}=\delta_{L,K}$ and
$J_0$ is the $34$-component vector of number
operators $J_0^{(L)}$. In terms of the mode expansion given in Table 1
we have $J_0^{(L)} = N_{(L)}$ except for the superghosts where
$J_0^{({\rm superghost})} = - N_{({\rm superghost})}$.
Finally, the $34 \times 34$ matrix $Y^E$
has all elements in the diagonal and the upper
triangle equal to zero, that is $Y_{KL}^E = 0$ for $K \leq L$, while
$Y_{KL}^E = \pm 1$ for $K > L$. It is at this point, i.e. in the
definition
of the $Y$ matrix, that we make use of the
ordering chosen for the fermions: The fermion $\Psi_{(L)}$ carries a
cocycle factor involving the number operator $J_0^{(K)}$ if and only if
$L > K$.

(The cocycle factor $C^{(L)}_{\rm gh}$ involves the number operators of
the reparametrization ghosts and of the $(\eta,\xi)$-system and is given
explicitly in ref.~[\Ref{ammedm}]. For the present purposes it is
sufficient to notice that
$ (C_{\rm gh}^{(L)})^{\dagger} = C_{\rm gh}^{(L)} = (C_{\rm gh}^{(L)})^{-1} $
where the first equality sign holds if we exclude the $\xi$ zero mode
and the second is a triviality, since the operator only takes values
$\pm 1$ on any string state.)

Although it seems that the choice of ordering is totally arbitrary, this
is not so. Indeed, the anomalous behaviour
$$ \left( J_0^{({\rm superghost})} \right)^{\dagger} = -
J_0^{({\rm superghost})} - 2 \efr
under hermitean conjugation forces us {\it to assign the highest number to
the superghost-related fermions}, i.e. $L=34$,
so that $J_0^{(34)}$ never appears
in the cocycle operators. This is because
to be consistent with the fact that
$\Psi_{(L)}^- = (\Psi_{(L)}^+)^*$ for $ L=1,\ldots,33$, we must have
$(C_{(L)})^{\dagger} = (C_{(L)})^{-1}$.

Notice, however,
that the superghost bosonization formula is not consistent with
hermitean conjugation, since $\Psi_{(34)}^{\pm}$ as
well as $e^{\pm \Phi_{(34)}}$ are hermitean operator fields, but
$(C_{(34)})^{\dagger} = (C_{(34)})^{-1}$. There seems to be
no way to avoid this problem in the Euclidean formulation, but we will see
that it is removed when we reformulate the bosonization in Minkowski
space-time.

\sjump

{\bf 4.} In Minkowski space-time, as it is clear from Table 1, the fermions
$\psi^{\pm}_{(0)}$ are hermitean and they now play a role
quite analogous to that of the superghost-related fermions
$\Psi^{\pm}_{(34)}$. The number operator $N_{(0)}$
is anti-hermitean, rather than hermitean, and therefore, if the bosonization
formulae for the $32$ complex fermions beside $\psi_{(0)}^{\pm}$
are to retain the correct hermiticity
properties, we must arrange for their
cocycle operators to depend neither on
$J_0^{(34)}$ nor on the number operator corresponding to
$\psi_{(0)}^{\pm}$. To ensure this, {\it in the Minkowski case we always
choose an ordering for the fermions such that $\psi_{(0)}^\pm$ is
given the number\/} $L=33$, retaining $L=34$ for $\psi_{(11)}^{\pm}$.
A natural choice for such an ordering is
$(\bar{1},\ldots,\overline{22};2,\ldots,10;1,0;11)$.

As long as we do not consider
hermitean conjugation, the only change as compared to the Euclidean case
is a reshuffle in the ordering of the fermions.
In particular it follows that the consistency conditions discussed in
ref.~[\Ref{ammedm}] ---which constrain the choices of the 561 independent
signs in the $Y$ matrix--- are exactly the same in the Euclidean and
Minkowski case.~\note{The two matrices $Y^E$ and $Y^M$ are indeed
related just by an interchange of rows and columns in the matrix
$\tilde{Y}$ introduced in ref. [\Ref{ammedm}] (see also below).}

The last question that remains to be addressed is whether the
bosonization formula is also consistent with hermitean conjugation in
the cases $L=33$ and $L=34$. Since $\Psi^\pm_{(K)}$ and $e^{\pm \Phi_{(K)}}$
are hermitean for $K=33,34$, we would like $C_{(33)}$ and
$C_{(34)}$ to be hermitean too. A priori this does not seem to be the
case, since
$$ \left( C_{(K)} \right)^{\dagger}  =
C_{(K)} \exp \left\{ -2\pi i \sum_{L=1}^{32} Y_{KL}^M
J_0^{(L)} \right\} \equiv C_{(K)} F_{(K)} \qquad {\rm for} \qquad
K=33,34 \nfr{cocycleherm}
However, when acting on any string state in the theory, the factor
$F_{(K)}$ appearing in eq.~\cocycleherm\ is
effectively equal to one. It is sufficient to check this on the ground
states since any raising operator has $J_0^{(L)} = $ integer.
A generic ground state is created by a vertex operator having the form
$$ {\cal V}_{\Bbb A} (z,\bar{z}) = S_{\Bbb A} (z,\bar{z}) e^{i k \cdot
X(z,\bar{z})} \ , \nfr{groundstate}
where $S_{\Bbb A} \equiv \prod_{L} S^{(L)}_{{\Bbb A}_{L}}$ and
$S^{(L)}_{{\Bbb A}_{L}} = \exp \{{\Bbb A}_{L} \Phi_{(L)}\}
(C_{(L)})^{{\Bbb A}_L}$. Acting on such a ground state, $F_{(K)}$ assumes
the eigenvalue
$$\eqalignno{ & F_{(K)} =
\exp \left\{ -2\pi i \sum_{L=1}^{32} Y_{KL}^M {\Bbb A}_L
\right\} = & \nameali{effectivelyone} \cr
&\qquad\quad\exp\left\{ -2\pi i \left( \varphi_K[{\Bbb A}] -
\tilde{Y}_{K,33}^M
{\Bbb A}_{33} - \tilde{Y}_{K,34}^M {\Bbb A}_{34} \right) \right\}
\qquad {\rm for} \ K=33,34 \cr } $$
where
$$ \varphi_K [{\Bbb A}] \equiv \sum_{L=1}^{34} \tilde{Y}_{KL}^M {\Bbb A}_L
\ {\rm mod} \ 2 \efr
and the matrix $\tilde{Y}_{KL}^M$  is defined by
$ \tilde{Y}_{KL}^M = Y_{KL}^M$ for $K > L$ and
$ \tilde{Y}_{KL}^M = -Y_{LK}^M$ for $K < L$,
while for $L=K$ we take $\tilde{Y}_{LL}^M = \pm \epsilon$, choosing
$+\epsilon$ if $L$ corresponds to one of the fermions $\psi_{(l)}$,
$l=0,1,\ldots,10$, and $-\epsilon$ otherwise. Here
$\epsilon = \pm 1$ keeps track of the phase encountered in the OPE
$S^{(L)}_{{\Bbb A}_L} (z,\bar{z}) S^{(L)}_{{\Bbb B}_L} (w,\bar{w})$ when
writing $(z-w) = e^{i \epsilon \pi} (w-z)$.
The quantity $\varphi_{K} [{\Bbb A}]$ keeps track of the statistics of
the operator $S_{\Bbb A}$, as encoded in the formula
$$ S^{(L)}_{{\Bbb B}_L} (z,\bar{z}) S_{\Bbb A} (w,\bar{w}) =
S_{\Bbb A} (w,\bar{w}) S^{(L)}_{{\Bbb B}_L} (z,\bar{z}) e^{i \pi {\Bbb
B}_L \varphi_L [{\Bbb A}]}\ . \nfr{statistics}

As was shown in ref.~[\Ref{ammedm}], one of the consistency conditions on
$Y^M$ ---following from the requirement that the picture changing
operator should always satisfy Bose statistics--- is that
$$ \varphi_{34}[{\Bbb A}] \eqmodtwo \varphi_{33}[{\Bbb A}] = {\rm integer}
\efr
for all vertex operators of the type \groundstate\ existing in the
theory. Furthermore,
world-sheet supersymmetry implies that either ${\Bbb A}_{33}$ and
${\Bbb A}_{34}$ are both integer (if the sector describes space-time
bosons) or both half-integer (if the sector describes space-time fermions).
It thus follows that the factor $F_{(K)}$ in eqs.~\cocycleherm,
\effectivelyone\ is effectively equal to one.

Notice that this argument, ensuring well-defined hermiticity properties
of the bosonized expression for
$\Psi^{\pm}_{(34)}$, does not work in the Euclidean case. There
all $33$ fermion number operators appearing in $C_{(34)}$ are
hermitean, and therefore one finds instead of \cocycleherm
$$ \left( C_{(34)} \right)^{\dagger} =
C_{(34)} \exp \left\{ -2\pi i \sum_{L=1}^{33} Y_{34,L}^E
J_0^{(L)} \right\} \qquad ({\rm Euclidean \ case})\ . \efr
By repeating the argument above we see that $C_{(34)}$ is actually
hermitean between states with integer value of ${\Bbb A}_{34}$ but
anti-hermitean between states with half-integer values of ${\Bbb A}_{34}$.
{}From this point of view, bosonization seems more well-defined in the
Minkowski case.
\sjump

{\bf 5.} We have presented a prescription for bosonizing the free
world-sheet fermions of a string theory embedded in Minkowski
space-time.

To summarize, we have seen that the bosonization in Minkowski
space-time differs from the one in Euclidean space-time in two ways.
First, the world-sheet fermions $\psi_{(0)}^{\pm}$, which are related to
the time direction in space-time, are assigned the label
$L=33$, whereas in the Euclidean case they could be assigned any value
$L \in \{1,\ldots,33\}$. This reordering of the fermions implies
that the explicit representation of the gamma matrices, as well as other
group theoretical objects, could be different in the Minkowski
and Euclidean formulation. But as long as all quantities satisfy the
correct group theoretical properties, as ensured by the cocycle
consistency conditions~[\Ref{ammedm}],
the final result in the computation of an amplitude should be
independent of what representations we happen to be using.

More important, the hermiticity properties of the fields
$\psi_{(0)}^{\pm}$ are different in the Euclidean and Minkowski
formulations. This means that the map between
$\vert {\rm ``in"} \rangle$ and
$\langle {\rm ``out"} \vert$ states is different in the two cases.
This should not come as a surprise, since it is well known
that whereas the spinor representation is unitary in the Euclidean case
it is not in the Minkowski case.

The bosonization prescription allows us to compute string
scattering amplitudes without the need to make a rotation to Euclidean
space-time. In string models based on free fields, such as the KLT
models, all correlation functions involved in the computation
can be evaluated explicitly.
As a simple example, and a check of our prescription,
we have considered the one-loop 3-point amplitude that was computed in
Euclidean space-time in ref.~[\Ref{ammedm}] and we have redone this
computation entirely in Minkowski space-time. As expected,
the result agrees with what we obtain by just Wick rotating the
Euclidean result, although the explicit form of the gamma matrices,
the mass matrix and the generalized charge conjugation matrix turns out
to be different in the two cases, as does the precise relation between
the vertex operators describing incoming and outgoing ``electrons''.
\references
\beginref
\Rref{GSW}{M.B.~Green, J.H.~Schwarz and E.~Witten, {\it Superstring
Theory}, Cambridge University Press, 1987.}
\Rref{Sonoda}{H.~Sonoda, Nucl.Phys. {\bf B326} (1989) 135.}
\Rref{KLT}{H.~Kawai, D.C.~Lewellen and S.-H.H.~Tye, Nucl.Phys.
{\bf B288} (1987) 1.}
\Rref{Koste}{V.A.~Kostelecky, O.~Lechtenfeld, W.~Lerche, S.~Samuel and
S.~Watamura, Nucl.Phys. {\bf B288} (1987) 173.}
\Rref{PDV1}{P.~Di Vecchia, M.L.~Frau, K.~Hornfeck, A.~Lerda, F.~Pezzella
and S.~Sciuto, Nucl.Phys. {\bf B322} (1989) 317.}
\Rref{ammedm}{A.~Pasquinucci and K.~Roland, ``{\sl On the computation
of one-loop amplitudes with external fermions in 4d heterotic
superstrings\/}'', preprint NBI-HE-94-47, hep-th/9411015, Nuclear Physics
{\bf B} in press.}
\Rref{DP}{E.~D'Hoker and D.H.~Phong, ``{\sl The box graph in superstring
theory\/}'', preprint Columbia/UCLA/94/TEP/39, hep-th/9410152.}
\endref
\ciao
